\documentclass[onecolumn,preprintnumbers,eqsecnum,12pt]{revtex4}
\usepackage{amsfonts}
\usepackage{amsmath}
\usepackage{epsf}
\usepackage{graphicx}
\usepackage{dcolumn}
\usepackage{bm}

\begin{document}

\title{Generalized Vaidya Spacetime in Lovelock Gravity and
Thermodynamics on Apparent Horizon}

\author{Rong-Gen Cai$^{a,}$\footnote{e-mail address:
cairg@itp.ac.cn}, Li-Ming Cao$^{b,}$\footnote{e-mail address:
caolm@apctp.org}, Ya-Peng Hu$^{a,c,}$\footnote{e-mail address:
yapenghu@itp.ac.cn}, Sang Pyo Kim$^{d,b}$\footnote{e-mail address:
sangkim@kunsan.ac.kr}}

\address{$^{a}$
Institute of Theoretical Physics, Chinese Academy of Sciences, P.O.
Box 2735, Beijing 100190, China}
\address{$^{b}$
Asia Pacific Center for Theoretical Physics, Pohang, Gyeongbuk
790-784, Korea}
\address{$^{c}$
Graduate School of the Chinese Academy of Sciences, Beijing 100039,
China}
\address{$^{d}$Department of Physics, Kunsan National University,
Kunsan 573-701, Korea}

\vspace*{2.cm}
\begin{abstract}
We present a kind of generalized Vaidya  solutions in a generic
Lovelock gravity. This solution generalizes the simple case in
Gauss-Bonnet gravity reported recently by some authors. We study the
thermodynamics of apparent horizon in this generalized Vaidya
spacetime. Treating those terms except for the Einstein tensor as an
effective energy-momentum tensor in the gravitational field
equations, and using the unified first law in Einstein gravity
theory, we obtain an entropy expression for the apparent horizon. We
also obtain an energy expression of this spacetime, which coincides
with the generalized Misner-Sharp energy proposed by Maeda and
Nozawa in Lovelock gravity.
\end{abstract}

\maketitle

\newpage

\section{Introduction}

The Lovelock gravity~\cite{Love} is a natural generalization of
general relativity (Einstein gravity theory) in higher dimensional
spacetimes. Its action is a sum of some dimensionally extended Euler
densities, where there are no more than second order derivatives
with respect to metric in equations of motion. Also it is known that
the Lovelock gravity is free of ghost. Over the past years, due to
the development of some theories in higher dimensional ($>4$)
spacetimes such as string theories, brane world scenarios etc, the
Lovelock gravity has attracted a lot of attention, and some new
features and properties have been revealed.

The action of Lovelock gravity can be written as
\begin{equation}  \label{1eq1}
\mathcal{L} = \sum^p_{i=0}c_i \mathcal{L}_i,
\end{equation}
where $p\leq [(n-1)/2]$($[N]$ denotes the integer part of the number
$N$), $c_i$ are arbitrary constants with dimension of
$[\mathrm{Length}]^{2i-2}$, $n$ is the spacetime dimension and
$\mathcal{L}_i$ are the Euler densities
\begin{equation}  \label{Eulerdensities}
\mathcal{L}_i=\frac{1}{2^i}\sqrt{-g}\delta^{a_1\cdots
a_ib_1\cdots b_i}_{c_1\cdots c_id_1\cdots d_i}R^{c_1d_1}_{~~~~
a_1b_1}\cdots R^{c_id_i}_{~~~~ a_ib_i},
\end{equation}
where the generalized delta function is totally antisymmetric in
both sets of indices. From the Lagrangian~(\ref{1eq1}), one can get
the equations of motion, $\mathcal{G}_{ab}=0$, where
\begin{equation}
\mathcal{G}^{a}_{b}=\sum_{i=0}^{p}\frac{1}{2^{i+1}} c_{i}\delta^{aa_1%
\cdots a_ib_1\cdots b_i}_{bc_1\cdots c_id_1\cdots
d_i}R^{c_1d_1}_{~~~~ a_1b_1}\cdots R^{c_id_i}_{~~~~ a_ib_i}\, .
\label{eom}
\end{equation}
After introducing matter into the theory, we can get the equations
of motion including the energy-momentum tensor of matter
\begin{equation}  \label{eomwithmatter}
\mathcal{G}_{ab}=8\pi G T_{ab}\, .
\end{equation}
%Here and in equation (\ref{eom}), the symbol $\delta^{a_1\cdots
%a_ib_1\cdots b_i}_{c_1\cdots c_1d_1\cdots d_i}$ is Generalized
%Kronecher-Delta.
From~(\ref{eom}), it can be shown that the
equations of motion do not have more than second order derivatives
with respect to metric and that the Lovelock theory has the same
degrees of freedom as the ordinary Einstein gravity theory. Just
because of this fact, the Lovelock gravity is free of ghost when
expanded on a flat spacetime, avoiding any problem with
unitarity~\cite{Zwie}. Note that, $\mathcal{L}_0$ denotes the unity,
and $\mathcal{L}_1$ gives us the usual curvature scalar term, while
$\mathcal{L}_2$ is just the Gauss-Bonnet term. Usually in order for
the Einstein gravity to be recovered in the low energy limit, the
constant $c_0$ should be identified as the cosmological constant up
to a constant and $c_1$ should be positive (for simplicity one may
take $ c_1=1$). For example, in $n=4$, we have
\begin{equation}
\frac{1}{2}c_0 g_{ab}+c_1(R_{ab}-\frac{1}{2}Rg_{ab})=8\pi G T_{ab}\,
.
\end{equation}
This is just the Einstein equation with the cosmological constant
$\Lambda=-c_0/2$ if we set $c_1$ to be one. When $n=5$ and $p=2$,
after expanding the Kroneker-Delta, one gets the equations of motion
for Gauss-Bonnet gravity in the usual manner.

In the literature on the Lovelock gravity, the most extensively
studied theory is the so-called Einstein-Gauss-Bonnet (EGB) gravity.
The EGB gravity is a special case of Lovelock's theory of
gravitation, whose Lagrangian just contains the first three terms in
(\ref{1eq1}). The Gauss-Bonnet term naturally appears in the low
energy effective action of heterotic sting
theory~\cite{Lowenergylimit}. Spherically symmetric black hole
solutions in the Gauss-Bonnet gravity have been found and discussed
in \cite{Des,Whee,Myers}, and topological nontrivial black holes
have been studied in \cite{Cai1}. Rotating Gauss-Bonnet black holes
have been discussed in \cite{rotating}. Some other extensions such
as including the perturbative AdS black hole solutions in gravity
theories with second order curvature corrections could be seen
in~\cite{others}. In addition, the references in~\cite{Cai3} have
investigated the holographic properties associated with the
Gauss-Bonnet theory. And the papers
in~\cite{Kobayashi,H.Maeda1,Dominguez,SGGhosh} gave some exact
solutions for Vaidya-like solution in the Einstein-Gauss-Bonnet
gravity.

For a generic case, although the Lagrangian~(\ref{1eq1}) looks
complicated, some exact black hole solutions have been found and
their associated thermodynamics was investigated
in~\cite{Banados:1993ur,Crisostomo:2000bb,Cai:1998vy,Aros:2000ij,Cai4,GG}.
In the so-called third-order Lovelock gravity, that is, containing
the first four terms in (\ref{1eq1}), some exact solutions have been
found in \cite{third}. Furthermore, it is also known that those
higher derivative terms in the Lagrangian~(\ref{1eq1}) with positive
coefficients arise as higher order corrections in superstring
theories, and their cosmological implication has been
studied~\cite{MO}.

In this paper, we are mainly interested in dynamical black hole
solutions in Lovelock gravity by generalizing those discussions in
\cite{Kobayashi,H.Maeda1,Dominguez, SGGhosh}.  The organization of
the paper is as follows. In Sec.~II, we present a kind of
generalized Vaidya spacetime in the general Lovelock gravity
(\ref{1eq1}).  In Sec.~ III, we study  thermodynamics of apparent
horizon of the generalized Vaidya spacetime, and give corresponding
entropy expression associated with the apparent horizon. In
addition, we propose a way to obtain the generalized Misner-Sharp
energy in the Lovelock gravity. Section. IV. is devoted to
conclusion and discussion.

%%==========================================================

\section{Generalized Vaidya Solution in Lovelock theory}

In the four-dimensional general relativity, the  Vaidya spacetime is
a typical dynamical one. In this spacetime there exists a pure
radiation matter. The metric of the spacetime can be written as
\begin{equation}
ds^{2}=-f(r,v)dv^{2}+2dvdr+r^{2}d\Omega _{2}^{2}\,.
\end{equation}%
where $f=1-2m(v)/r$, and $d\Omega _{2}^{2}$ is the line element of a
two-dimensional unit sphere. In this spacetime, the apparent horizon
is given by $f=0$, or $r=2m(v)$.  The energy-momentum tensor for the
radiation matter in the spacetime is given by $T_{ab}=\mu
l_{a}l_{b}$, where $l_{a}=(1,0,0,0)$ in coordinates
$(v,r,\theta,\phi)$. The quantity $\mu $ is the energy density of
the radiation matter. Now, in an $n$-dimensional spacetime, assume a
similar form of spherically symmetric metric
\begin{equation}
ds^{2}=-f(r,v)dv^{2}+2dvdr+r^{2}d\Omega _{n-2}^{2}\,.
\label{VaidyaLovelock}
\end{equation}
The energy-momentum tensor of radiation matter has a similar form as
that in four dimensions. For this metric, it is not hard to
calculate the nonvanishing components of Riemann tensor given by
\begin{eqnarray}
&&R_{~~vr}^{vr}=-\frac{f^{\prime \prime }}{2}\,,\quad R_{~~vj}^{vi}=-\frac{%
f^{\prime }}{2r}\delta _{j}^{i}\,,\quad R_{~~rj}^{ri}=-\frac{f^{\prime }}{2r}%
\delta _{j}^{i}\, , \nonumber  \\
&&R_{~~vj}^{ri}=-\frac{\dot{f}}{2r}\delta _{j}^{i}\,,\quad
R_{~~rj}^{vi}=0\,,\quad R_{~~kl}^{ij}=\frac{1-f}{r^{2}}\delta _{kl}^{ij}\,.
\end{eqnarray}%
Here a prime/overdot denotes  the derivative with respect to $r/v$.
Substituting these results into the equations of motion
(\ref{eomwithmatter}), and using identities
\begin{equation*}
\delta _{b_{1}\cdots b_{m}}^{a_{1}\cdots a_{m}}\delta
_{a_{m}}^{b_{m}}=[n-(m-1)]\delta _{b_{1}\cdots b_{m-1}}^{a_{1}\cdots a_{m-1}}
\end{equation*}%
and
\begin{equation*}
\delta _{b_{1}\cdots b_{m-1}b_{m}}^{a_{1}\cdots a_{m-1}a_{m}}\delta
_{a_{m-1}a_{m}}^{b_{m-1}b_{m}}=2[n-(m-1)][n-(m-2)]\delta _{b_{1}\cdots
b_{m-2}}^{a_{1}\cdots a_{m-2}},
\end{equation*}%
we find the  equations of motion
\begin{equation}
\mathcal{G}_{v}^{v}=\sum_{i}^{p}c_{i}\frac{(n-2)!}{(n-2i-1)!}\left[ i\left( -%
\frac{f^{\prime }}{2r}\right) \left( \frac{1-f}{r^{2}}\right) ^{i-1}+\frac{1%
}{2}(n-2i-1)\left( \frac{1-f}{r^{2}}\right) ^{i}\right] =0\,,  \label{Gvv}
\end{equation}%
\begin{equation}
\mathcal{G}_{v}^{r}=\sum_{i}^{p}c_{i}\frac{(n-2)!}{(n-2i-1)!}\left[ i\left( -%
\frac{\dot{f}}{2r}\right) \left( \frac{1-f}{r^{2}}\right)
^{i-1}\right] =8\pi G\mu \, ,  \label{Grv}
\end{equation}
and
\begin{eqnarray}
\label{Gii} \mathcal{G}^j_k&=& 0 =\delta^j_k \sum_{i}^p
c_i\frac{1}{2}\Bigg{\{}
\frac{(n-3)!}{(n-2i-3)!}\left(\frac{1-f}{r^2}\right)^{i}\nonumber
\\
&+&4i \frac{(n-3)!}{(n-2i-2)!} \left(-\frac{f^{'}}{2r}\right)
\left(\frac{1-f}{r^2}\right)^{i-1}
\nonumber \\
&+&2i \frac{(n-3)!}{(n-2i-1)!}
\left(-\frac{f^{''}}{2}\right)\left(\frac{1-f}{r^2}\right)^{i-1}
\nonumber
\\
&+&4i(i-1)\frac{(n-3)!}{(n-2i-1)!}\left(-\frac{f^{'}}{2r}\right)^2
\left(\frac{1-f}{r^2}\right)^{i-2}\Bigg{\}}\, .
\end{eqnarray}
Other components of $\mathcal{G}^a_b$ are given by
$\mathcal{G}^r_r=\mathcal{G}^v_v$, $\mathcal{G}_r^v=0$. The
components $\mathcal{G}_j ^i$ are not independent, because they are
linearly expressed in terms of $\partial_r\mathcal{G}_v^v$ and
$\mathcal{G}_v^v$, as will be shown below.  Defining a new function
$F(v,r)$
\begin{equation}
F(v,r)=\frac{1-f(v,r)}{r^{2}}\,,
\end{equation}%
we can put the equations (\ref{Gvv}) and (\ref{Gii}) into the forms
\begin{equation}
\sum_{i}^{p}c_{i}\frac{(n-2)!}{2(n-2i-1)!}\frac{1}{r^{n-2}}\left[
r^{n-1}F^{i}\right] ^{\prime }=0\,,  \label{Fequation}
\end{equation}%
and
\begin{equation}
\label{Giicompact}
\mathcal{G}^j_k=\delta^j_k\sum_i^pc_i\frac{(n-3)!}{2(n-2i-1)!}\frac{1}{r^{n-3}}\left[r^{n-1}F^i\right]''=0\,
.
\end{equation}
>From these two equations, it is easy to show
$\mathcal{G}^i_j=\delta^i_j\left[r\partial_r\mathcal{G}_v^v
/(n-2)+\mathcal{G}_v^v\right]$, so $\mathcal{G}^i_j=0$ do not
yield independent equations. Integrating the equation (\ref{Fequation})
leads to an order-$p$ algebraic equation for $F(v,r)$ or $%
f(v,r)$
\begin{equation}
\sum_{i}^{p}c_{i}\frac{(n-2)!}{(n-2i-1)!}F^{i}=\frac{16\pi
Gm(v)}{\Omega _{n-2}r^{n-1}}\,,  \label{massandF}
\end{equation}%
where $m(v)$ is an arbitrary function of $v$, which appears as an
integration constant. (Certainly, to ensure some energy condition,
this mass function should be positive.) The coefficient $16\pi
G/\Omega _{n-2}$ is chosen  such that $m$ can be interpreted as the
mass of the solution when $m$ is a constant. Using
\begin{equation}
\label{realtion1}
\frac{\partial }{\partial v}F^{i}=iF^{i-1}\dot{F}=i\left( -\frac{\dot{f}}{%
r^{2}}\right) \left( \frac{1-f}{r^{2}}\right) ^{i-1}\,,
\end{equation}
in the equation (\ref{Grv}), we have
\begin{equation}
8\pi G\mu =\sum_{i}^{p}c_{i}\frac{(n-2)!r}{2(n-2i-1)!}\frac{\partial }{%
\partial v}F^{i}\,.  \label{massEq}
\end{equation}%
Comparing equations (\ref{massandF}) and (\ref{massEq}), we obtain
\begin{equation}
\mu =\frac{\dot{m}(v)}{\Omega _{n-2}r^{n-2}}\,.  \label{energyareadensity}
\end{equation}%
Thus, we obtain a kind of radiating Vaidya spacetime in a generic
Lovelock gravity by solving  equation (\ref{massandF}). When the
dimension of spacetime is five, that is, $n=5$, the solution reduces
to the one reported by some authors in
references~\cite{Kobayashi,H.Maeda1,Dominguez, SGGhosh}, while in
$n=4$, the solution is just the familiar Vaidya solution of general
relativity.

Now we further generalize the Vaidya spacetime in Lovelock gravity
to more general case.  Note that for the metric
(\ref{VaidyaLovelock}) we have $\mathcal{G}^r_r=\mathcal{G}^v_v$, so
the energy-momentum tensor of matter has to satisfy $T^r_r=T^v_v$.
Certainly, the matter of pure radiation discussed above satisfies
the constraint. In fact, they are $T^r_r=T^v_v=0$. If we further
assume the spherical part of the energy-momentum tensor has the form
$T^{i}_{i}=\sigma T^r_r=\sigma T^v_v$ (where $\sigma$ is a constant,
and the repeat index $i$ does not sum), then from the equation
$\nabla_aT^{a}_b=0$ or the explicit expressions of $\mathcal{G}^a_b$
in equations (\ref{Gvv}), (\ref{Grv}) and (\ref{Gii}), we can find
\begin{equation}
\label{TRVEq}
\partial_v T^v_v+\partial_rT^r_v+\frac{n-2}{r}T^r_v=0\, ,
\end{equation}
and
\begin{equation}
\label{TRREq}
\partial_r T^r_r+\frac{(n-2)(1-\sigma)}{r}T^r_r=0\, .
\end{equation}
As a result, for the pure radiation matter with $T^r_r=T^v_v=0$, one
can find that $T^r_v$ has to be proportional to $1/r^{n-2}$. This is
consistent with the equation (\ref{energyareadensity}). Next, in the
case with $T^r_r=T^v_v\ne 0$, the equation (\ref{TRREq}) tells us
that $T^r_r$ and $T^v_v$ have the form
\begin{equation}
T^r_r=T^v_v=\mathcal{C}(v) r^{-(n-2)(1-\sigma)}\, ,
\end{equation}
where $\mathcal{C}(v)$ is a function of $v$. The off-diagonal part
of the energy-momentum tensor $T^a_b$, i.e., the component $T^r_v$
has to satisfy the equation (\ref{TRVEq}). Now the equations of
motion $\mathcal{G}^v_v = \mathcal{G}^r_r =  8 \pi G \mathcal{C}(v)
r^{-(n-2)(1-\sigma)}$ modify the equation (\ref{Fequation}) to
\begin{equation}
\sum_{i}^{p}c_{i}\frac{(n-2)!}{2(n-2i-1)!}\left[ r^{n-1}F^{i}\right]
^{\prime }=8\pi G~\mathcal{C}(v) r^{(n-2)\sigma}\,.
\label{Fequationgeneral}
\end{equation}
Integrating this equation, we have
\begin{equation}
\label{moregeneralsolution}
\sum_{i}^{p}c_{i}\frac{(n-2)!}{(n-2i-1)!}F^{i}=16\pi G\left(\frac{
m(v)}{\Omega_{n-2} r^{n-1}}+ \frac{\mathcal{C}(v)\Theta
(r)}{r^{n-1}}\right)\, ,
\end{equation}
where $m(v)$ is an arbitrary function, appearing as an integration
constant again. Here, $\Theta(r) = \int dr r^{(n-2)\sigma}$, which
 is
\begin{equation}
\Theta(r)= \mathrm{ln}
(r)\, ,
\end{equation}
for $\sigma=-1/(n-2)$ and
\begin{equation}
\Theta(r)= \frac{r^{(n-2)\sigma+1}}{(n-2)\sigma+1} \, ,
\end{equation}
otherwise. Certainly, to ensure some energy condition for the
energy-momentum tensor, the parameter $\sigma$ and function $m(v)$
and $\mathcal{C}(v)$ should satisfy certain consistency relations.
For example, $\mathcal{C}(v)\le 0$ and $-1\le \sigma \le 0$. These
relations have been discussed in \cite{Dominguez}. One can see from
the relation (\ref{realtion1}) that $T^r_v$ is given by the partial
derivative of the equation (\ref{moregeneralsolution}) with respect
to $v$. Thus we have
\begin{equation}
\label{mutilde} T^r_v=\tilde{\mu}=\frac{\dot{m}(v)}{\Omega_{n-2}
r^{n-2}}+\frac{\dot{\mathcal{C}}(v)\Theta (r)}{r^{n-2}}\, .
\end{equation}
This is consistent with the equation (\ref{TRVEq}). So the
energy-momentum tensor of matter in this case can be written as
\begin{equation}
\label{energymomentumtensor} T_{ab}=\tilde{\mu}
l_al_b-P(l_an_b+n_al_b)+\sigma P q_{ab}\, ,
\end{equation}
where $n_a$ is a null vector which satisfies $l_an^a=-1$. In
coordinates $\{v,r,\cdots\}$, we have $l_a=(1,0,0,\cdots)$, and
$n_a=(f/2,-1,0,\cdots)$. The tensor $q_{ab}$ is a projection
operator which is given by $q_{ab}=g_{ab}+l_an_b+l_bn_a$. So the
metric (\ref{VaidyaLovelock}) can be put into the form
$g_{ab}=h_{ab}+q_{ab}$, where
\begin{equation}
h_{ab}=-l_an_b-l_bn_a
\end{equation}
is the metric of the two-dimensional spacetime transverse to the
$(n-2)$-dimensional sphere. Certainly, in  coordinates
$\{v,r,\cdots\}$, the line element of $h_{ab}$ can be expressed as $
-f(v,r)dv^2+2dvdr$. The quantity $P$ is the radial pressure with the
form $P=-\mathcal{C}(v)r^{-(n-2)(1-\sigma)}$.

Generally, it is not easy to solve equation
(\ref{moregeneralsolution}) when $p$ is greater than one. However,
for some special cases, one can explicitly get analytic solutions.
Now, we give some simple examples.

\indent (1). Gauss-Bonnet theory: The simple radiating Vaidya
solution in the Gauss-Bonnet gravity without a cosmological constant
has been found in the references~\cite{Kobayashi,H.Maeda1,Dominguez,
SGGhosh}. Here we can directly solve the second order algebraic
equation (\ref{moregeneralsolution}), and give the solution
\begin{equation}
ds_{GB}^{2}=-f(v,r)d\upsilon ^{2}+2dvdr+r^{2}d\Omega _{n-2}^{2}\, ,
\label{GBVaidya}
\end{equation}
where $f(v,r)$ satisfies the equation (\ref{moregeneralsolution}),
which in this case  becomes
\begin{equation}
F+\alpha (n-3)(n-4)F^2=\frac{16  \pi G}{n-2}\left(\frac{
m(v)}{\Omega_{n-2} r^{n-1}}+ \frac{\mathcal{C}(v)\Theta
(r)}{r^{n-1}}\right)\, .
\end{equation}
Here, we have set $c_0=0$, $c_1=1$ and $c_2=\alpha$. This parameter
$\alpha$, called the Gauss-Bonnet coefficient, has the dimension of
length squared. Since it is an algebraic equation of order two, the
solutions have two branches in general. Only one branch has the
general relativity limit and is given by
\begin{equation}
f(v,r)=1+\frac{r^{2}}{2(n-3)(n-4)\alpha
}\left[1-\sqrt{1+\frac{64\pi(n-3)(n-4)\alpha}{n-2}\left(\frac{
m(v)}{\Omega_{n-2}r^{n-1}}+\frac{\mathcal{C}(v)\Theta(r)}{r^{n-1}}\right)}\right]\,
,
\end{equation}
where, for simplicity, we have set $G=1$. From $f(v,r)=0$, we can
obtain the trapping horizon, which is also its apparent horizon in
this case. The radius of apparent horizon satisfies
\begin{equation}
\label{apparent horizon} r_{A}^{n-3}+\alpha(n-3)(n-4)
~r_{A}^{n-5}=\frac{16\pi}{n-2}\left(\frac{m(v)}{\Omega_{n-2}}+\mathcal{C}(v)\Theta(r_A)\right)\,
.
\end{equation}
Thus, in general, the radius of the apparent horizon is a function
of $v$. This solution is the same as the one in \cite{Dominguez} if
a cosmological constant is included.

\indent (2). Dimensionally continued Lovelock gravity: An
interesting case is to choose some special values for the
coefficients $c_i$, $i=0,\cdots , p$. In~\cite{Banados:1993ur,
Crisostomo:2000bb, Aros:2000ij, Cai:1998vy} a set of special
coefficients has been chosen so that the equation
(\ref{moregeneralsolution}) has a simple expression. In odd
dimensions the action is the Chern-Simons form for the AdS group,
while in even dimensions it is called Born-Infeld theories
constructed with the Lorentz part of the AdS curvature tensor. In
the odd-dimensional case, say $n=2p+1$, i.e., Chern-Simons theory,
we can choose
\begin{equation}
c_i=(n-2i-1)!\binom{p}{i}\ell^{-n+2i}\, ,
\end{equation}
where the parameter $\ell$ is a length scale. Note that $c_i$ differ
from the coefficients $\alpha_i$ in the
references~\cite{Banados:1993ur,Crisostomo:2000bb,Cai:1998vy,Aros:2000ij}
only by a factor of $(n-2i)!$.  At the same time we choose $1/16\pi
G$ as
\begin{equation}
\frac{\Omega_{n-2}}{16\pi G}=\frac{\ell}{(n-2)!}\, .
\end{equation}
Then the equation (\ref{moregeneralsolution}) gives the solution
\begin{equation}
\label{chernsimons}
f(v,r)=1-\left[m(v)+\Omega_{n-2}\mathcal{C}(v)\Theta(r)\right]^{\frac{1}{p}}+\frac{r^2}{\ell
^2}\, ,
\end{equation}
For the even-dimensional case, say $n=2p+2$, we set $c_i$,
$i=1,\cdots, p$, to be
\begin{equation}
c_i=(n-2i-1)!\binom{p}{i}\ell^{-n+2i}\, ,
\end{equation}
and the gravity coupling constant
\begin{equation}
\frac{\Omega_{n-2}}{16\pi G}=\frac{\ell^2}{(n-2)!}\, ,
\end{equation}
Then the equation (\ref{moregeneralsolution}) gives the solution
\begin{equation}
\label{Born-Infeld}
f(v,r)=1-\left[\frac{m(v)+\Omega_{n-2}\mathcal{C}(v)\Theta(r)}{r}\right]^{\frac{1}{p}}+\frac{r^2}{\ell^2}\,
.
\end{equation}
The solutions (\ref{chernsimons}) and (\ref{Born-Infeld}) reduce to
the static cases if $m(v)$ and $\mathcal{C}(v)$ do not depend on
coordinate $v$. In these cases, the solutions have a unique AdS
vacuum~\cite{Banados:1993ur,Crisostomo:2000bb,Cai:1998vy,Aros:2000ij}.
These solutions (\ref{chernsimons}) and (\ref{Born-Infeld}) with
vanishing $\mathcal{C}(v)$ have already been found by M. Nozawa and
H. Maeda in~\cite{Nozawa:2005uy}.

\indent (3). Pure Lovelock gravity: In this
theory~\cite{Kastor:2006vw, Giribet:2006ec, Cai:2006pq}, only two of
coefficients $c_i$ are non-vanishing: one is $c_0$ and the other is
$c_k$ with $1\le k\le p$. We can normalize them as
$c_0/(n-1)(n-2)=-1/\ell^2$ and $c_k(n-1)!/(n-2k-1)!=\alpha^{2k-2}$,
where $\ell$ and $\alpha$ are two length scales. Then the equation
(\ref{moregeneralsolution}) becomes
\begin{equation}
\label{purelovelock} \alpha^{2k-2}F^k=\frac{1}{\ell^2}+\frac{16  \pi
G}{n-2}\left(\frac{ m(v)}{\Omega_{n-2} r^{n-1}}+
\frac{\mathcal{C}(v)\Theta (r)}{r^{n-1}}\right)\, .
\end{equation}
Thus, in the case of even dimensions, if the right hand of
(\ref{purelovelock}) is non-negative, we have
\begin{equation}
f(v,r)=1\pm\frac{r^2}{\alpha^2}\left[\frac{\alpha^2}{\ell^2}+\frac{16
\pi G\alpha^2}{n-2}\left(\frac{ m(v)}{\Omega_{n-2} r^{n-1}}+
\frac{\mathcal{C}(v)\Theta
(r)}{r^{n-1}}\right)\right]^{\frac{1}{k}}\, ,
\end{equation}
while, in the other case of odd dimensions,
\begin{equation}
f(v,r)=1+\frac{r^2}{\alpha^2}\left[\frac{\alpha^2}{\ell^2}+\frac{16
\pi G\alpha^2}{n-2}\left(\frac{ m(v)}{\Omega_{n-2} r^{n-1}}+
\frac{\mathcal{C}(v)\Theta
(r)}{r^{n-1}}\right)\right]^{\frac{1}{k}}\, .
\end{equation}
Thus, for $k=1$, one gets a generalized Vaidya black hole with a
cosmological constant. When $m(v)$ and $\mathcal{C}(v)$ are two
constants, these solutions  reduce to the static solutions.

For dynamical black holes, it is difficult to study their
thermodynamical properties. In four-dimensional Einstein gravity,
Hayward has proposed a method to discuss this issue~\cite{Hayward,
Hayward1, Hayward2, Hayward3}. In the next section, we will discuss
the thermodynamics of the apparent horizon of the new solutions in
this section.

\section{Entropy and energy of the apparent horizon}

In the early 1970s, it was found that four laws of black hole
mechanics in general relativity are very analogous to  four laws of
thermodynamics. Due to Hawking's discovery that black hole radiates
thermal radiation, it turns out that it is not just an analog, but
they are indeed identical with each other. Then thermodynamics of
black hole has been established soundly (for a review,
see~\cite{Wald:1999vt}). In black hole thermodynamics, the
temperature and entropy of a black hole are given by $T_{\rm
EH}=\kappa/2\pi$ and $S_{\rm EH}=A/4$, where $\kappa$ and $A$ are
the surface gravity and area of event horizon of the black hole,
respectively. In higher derivative theory of gravity, such as
Gauss-Bonnet gravity, it turns out that the area formula for black
hole entropy no longer holds. In fact, black hole entropy gets
corrections from these higher derivative terms. For a diffeomorphism
invariant theory, Wald~\cite{Noth} showed that the entropy of black
hole is a kind of Noether charge, and obtained a formula for black
hole entropy, now called Wald formula, associated with the event
horizon of black hole. For more details, see the review
 paper by Wald~\cite{Wald:1999vt}.

In the literature, most of discussions on thermodynamics of black
holes have been focused on stationary black holes. Thermodynamics of
a black hole is associated with the event horizon of the black hole,
which is the boundary of the past of future infinity. Therefore the
event horizon is a null hypersurface and depends on some global
structures of the spacetime, so it is difficult to study the event
horizon of a dynamical (time-dependent) spacetime. In general
relativity, there exists another kind of horizon, named apparent
horizon, which is  not heavily dependent of global structure of
spacetime. In the standard definition of apparent horizon, one first
slices the spacetime, and then finds the boundary of trapped region
in each slice.  This boundary (two-dimensional surface) is called
the apparent horizon~\cite{HawkingEllis}. Also the three-dimensional
hypersurface of the union of all these two-dimensional surfaces is
usually called the apparent horizon. Over the past years, several
generalizations of horizon have been proposed, such as the trapping
horizon by Hayward, the isolated horizon and the dynamical horizon
by Ashtekar {\it et al}. The relations and differences among those
horizons have been discussed in a recent
review~\cite{Ashtekar:2004cn}. For a dynamical black hole, the outer
trapping horizon and the dynamical horizon are not null
hypersurfaces but spacelike hypersurfaces of the
spacetime~\cite{Ashtekar:2004cn}. Therefore,  some well-known
results associated with the event horizon, such as Wald entropy
formula, may not be applicable to those horizons.

In this section, we will discuss the entropy and the energy
associated with the apparent horizon of the generalized Vaidya
solution in Lovelock gravity theory (\ref{moregeneralsolution}).
Before going on, let us first give a brief review on the work of
Hayward~\cite{Hayward, Hayward1, Hayward2, Hayward3}. Although his
work focuses on four-dimensional Einstein gravity, it can be
straightforwardly generalized to higher dimensional Einstein
gravity~\cite{Bak:1999hd, cai5}. These discussions reveal the deep
relation between equations of motion and thermodynamics of the
spacetimes. For relevant discussions, see
also~\cite{Jacobson:1995ab, Eling:2006aw, Padmanabhan:2002sha,
Paranjape:2006ca, Padmanabhan:2007en, Akbar:2008vz}. For an
$n$-dimensional spherically symmetric spacetime
$(\mathcal{M},g_{ab})$, we can write the metric in the double null
form
\begin{equation}
ds^2=h_{ab}dx^adx^b+r^2(x)d\Omega_{n-2}^2\, ,
\end{equation}
where $\{x^a\}$ are coordinates of the two-dimensional spacetime
$(M,h_{ab})$ which is transverse to the $(n-2)$-dimensional sphere,
$r(x)$ is the radius of the sphere. Similarly, one can also divide
the energy-momentum tensor of matter into two parts. One part
denoted by $T_{ab}$ (do not be confused with the total
energy-momentum tensor) corresponds to the two-dimensional spacetime
$h_{ab}$. From this energy-momentum tensor, one can define two
important physical quantities: the work density $W=-1/2
h^{ab}T_{ab}$, which corresponds to the work term in the first law,
and the energy supply $\Psi_a=T_{a}^{b}\partial _{b}r+W\partial
_{a}r$. By using these two quantities and the Misner-Sharp energy
inside the sphere with radius $r$,
\begin{equation}
E=\frac{(n-2)}{16\pi }\Omega _{n-2}r^{n-3}\left(1-h^{ab}\partial
_{a}r\partial _{b}r\right)\, , \label{Misner-Sharp}
\end{equation}
one can put some components of the Einstein equations into the
so-called unified first law
\begin{equation}
\label{unified first law} dE=A\Psi+WdV\, ,
\end{equation}
where $A=\Omega_{n-2}r^{n-2}$ and $V=\Omega _{n-2}r^{n-1}/(n-1)$ are
the area and the volume of an $(n-2)$-sphere with radius $r$. For a
vector $\xi$ tangent to the trapping horizon of the spacetime,
Hayward showed that on trapping horizon, one has~\cite{Hayward2}
\begin{equation}
\label{Clausius}
A\Psi_a\xi^a=\frac{\kappa}{8\pi}\mathcal{L}_{\xi}A=\frac{\kappa}{2\pi}\mathcal{L}_{\xi}S=T\delta
S\, ,
\end{equation}
where $S=A/4$, $T=\kappa/2\pi$, and $\kappa$ is the surface gravity
defined by $\kappa=D_aD^ar/2$. Here, $D_a$ is the covariant
derivative associated with metric $h_{ab}$. The $\delta$ operator in
the right hand side of the equation (\ref{Clausius}) should be
understood as follows: take a Lie derivative with respect to $\xi$,
and then evaluate it on the apparent horizon. Note that the left
hand side of the equation (\ref{Clausius}) represents the amount of
energy crossing the trapping horizon. Therefore the equation
(\ref{Clausius}) may be understood as the Clausius relation for
mechanics of dynamic black holes: $\delta Q=T\delta S$ with $\delta
Q=A\Psi_a\xi^a$. By projecting the unified first law onto the
trapping horizon, the first law of thermodynamics is given by
\begin{equation}
\delta E=T\delta S+W\delta V\, .
\end{equation}

In the generalized Vaidya spacetime found in the previous section,
the apparent horizon is just a kind of trapping horizon, so we do
not emphasize the difference between these two concepts in the
present paper. For the generalized Vaidya spacetime
(\ref{VaidyaLovelock}), the apparent horizon is given by $f(v,r)=0$.
From the definition of surface gravity, it is easy to find $\kappa=
f'(v,r_A)/2$.  We now discuss thermodynamics on apparent
horizon/trapping horizon for the dynamical solutions in Lovelock
gravity theory.  Following \cite{cai5}, we may move all terms of
$\mathcal{G}_{ab}$ except for the Einstein tensor into the right
hand side of the field equations $\mathcal{G}_{ab}=-8\pi T_{ab}$ and
rewrite them in the standard form for Einstein gravity
\begin{equation}
G_{ab}=8\pi T_{ab}=8\pi \left(T_{ab}^{(m)}+T_{ab}^{(e)}\right).
\label{Einsteinequation}
\end{equation}%
Here, $T_{ab}^{(m)}$ is the energy-momentum tensor of matter
(\ref{energymomentumtensor}). The effective energy-momentum tensor
$T_{ab}^{(e)}$ has the form
\begin{equation}
8\pi T_{ab}^{(e)}=H_{ab}=\mathcal{G}_{ab}-G_{ab} \,
,\label{Effective energy-momentum tensor}
\end{equation}
where $G_{ab}$ is the Einstein tensor of the spacetime. In this way,
we can go along the line of Einstein gravity, although we are
discussing a gravity theory beyond the Einstein gravity. Thus, the
work term and the energy supply defined on the two-dimensional
spacetime ($M^{2},h_{ab}$) are, respectively,
\begin{eqnarray}
W &=&-\frac{1}{2}h^{ab}T_{ab}=W^{(m)}+W^{(e)} \, ,\nonumber \\
\Psi _{a} &=&T_{a}^{b}\partial _{b}r+W\partial _{a}r=\Psi
_{a}^{(m)}+\Psi _{a}^{(e)}\, ,\label{WTandES}
\end{eqnarray}%
where $W^{(m)}$ and$\ \Psi _{a}^{(m)}$ are defined by using  $T_{ab}^{(m)}$, while $%
W^{(e)}$ and $\Psi _{a}^{(e)}$ by $T_{ab}^{(e)}$, and the
contraction is taken over the two-dimensional spacetime $h_{ab}$. It
is easy to find in our case,
\begin{equation}
W^{(m)}=-P=\mathcal{C}(v)r^{-(n-2)(1-\sigma)}\, ,
\end{equation}
so the energy supply for the matter is given by
\begin{equation}
\label{energysupplymatter} \Psi^{(m)}_a=\tilde {\mu}l_a\, .
\end{equation}
Now, because we are treating an effective Einstein gravity (i.e.
view the non-Einstein term $H_{ab}$ as an energy-momentum tensor of
higher curvature terms), the Misner-Sharp energy
(\ref{Misner-Sharp}) is applicable. For the generalized Vaidya
spacetime,  we can replace the term $h^{ab}\partial _{a}r\partial
_{b}r$ in equation (\ref{Misner-Sharp}) by $f(v,r)$. It is easy to
see that equations (\ref{WTandES}) and (\ref{Misner-Sharp}) satisfy
the unified first law (\ref{unified first law}).

On the apparent horizon, we have the Clausius-like equation
\begin{equation}
A\Psi_a \xi^a=A\Psi^{(m)}_a\xi^a+A\Psi^{(e)}_a\xi^a=\frac{\kappa
}{8\pi }\delta A  \, .\label{AHEnergysupplyForm}
\end{equation}
where $\xi$ is a vector field tangent to the apparent horizon. This
vector can be determined as follows. Since $f=0$ on the
$(n-2)$-dimensional marginal surface, the Lie derivative of $f$ with
respect to $\xi$ should vanish, so $\mathcal{L}_{\xi}f=0$ on this
surface. This means that on the apparent horizon we have
\begin{equation}
\xi^v\partial_vf+\xi^r\partial_rf=0\, .
\end{equation}
Noting that $\kappa=f'(v,r_A)/2$, we have
$\xi^r/\xi^v=-\dot{f}/2\kappa$. In the case of
$\dot{f}=\partial_vf=0$, we have $\xi^r=0$. Therefore on the
apparent horizon, it is reasonable to set $\xi_a$ to be
$$\xi_a=\left(\frac{\dot{f}}{2\kappa}\right)l_a+ n_a\, .$$
Here, the normalization of $\xi$ is not important and will not play
a key role in the following discussion.  It is easy to find
$\xi^a\xi_a=-\dot{f}/\kappa$. So the generator or the tangent vector
of the apparent horizon is not a null vector unless $\dot{f}=0$,
i.e., the static case. On the other hand, for the static cases, one
has $\xi^a=n^a=(\partial/\partial v)^a$ on the apparent horizon, as
expected.

Note that the heat flow $\delta Q$ is determined  by the pure matter
energy-momentum tensor or pure matter energy-supply. This point has
been emphasized in \cite{cai5}. Thus, on the apparent horizon we can
rewrite the equation (\ref{AHEnergysupplyForm}) as
\begin{equation}
\label{heatflowonAH} \delta Q\equiv A\xi^a\Psi_a^{(m)}=\frac{\kappa
}{8\pi }\delta A-A\xi^a\Psi_a^{(e)}\, .
\end{equation}
An interesting question is whether the right hand side of the
equation (\ref{heatflowonAH} ) can be cast into a form $T\delta S$
of the Clausius relation as in the Einstein gravity. The answer is
affirmative, since the right hand side of the (\ref{heatflowonAH})
indeed can be written in a form of $T\delta S$, where $S$ will be
given below. One can get this entropy expression by directly
substituting the expression of $H_{ab}$ in equation (\ref{Effective
energy-momentum tensor}) into the equation (\ref{heatflowonAH}). In
fact, the authors in \cite{cai5} obtain an entropy expression of
apparent horizon in Lovelock gravity, but in a setting of FRW
universe. Here let us stress that it is not always possible to
rewrite the right hand side of the equation (\ref{heatflowonAH}) in
a form $T\delta S$, scalar-tensor theory and $f(R)$ gravitational
theory being the counterexamples~\cite{cai5}.

In order to find the entropy expression on the apparent horizon,
here we use a little different method from the one in \cite{cai5}.
By using the explicit form of $\Psi_a^{(m)}$ in equation
(\ref{energysupplymatter}), we have
\begin{equation}
\label{heatflow} \delta Q\equiv A\Psi_a^{(m)}\xi^a=A\tilde
{\mu}l_a\xi^a=-A \tilde{\mu}\, .
\end{equation}%
>From equations (\ref{realtion1}), (\ref{moregeneralsolution}) and
(\ref{mutilde}), it is not hard to find
\begin{equation}
-2A\frac{\tilde{\mu}}{\dot{f}}=\frac{A}{8\pi}\sum_ic_i\frac{i(n-2)!}{(n-2i-1)!}r^{1-2i}\,
.
\end{equation}
Hence we have on the apparent horizon,
\begin{equation}
\label{deltaQeqTdeltaS} \delta Q=-A\tilde{\mu}=\frac{\kappa}{2\pi}
\left(\frac{\dot{f}}{2\kappa}\right)\left(\frac{A}{4}\sum_ic_i\frac{i(n-2)!}{(n-2i-1)!}r^{1-2i}\right)
=\frac{\kappa}{2\pi}\mathcal{L}_{\xi}
S =\frac{\kappa}{2\pi}\delta S\, ,
\end{equation}
where all the calculations are done on the apparent horizon. This is
very similar to the Clausius relation $\delta Q=T\delta S$ if we
define the temperature by $T=\kappa/2\pi$ and the entropy of
apparent horizon $S$ by
\begin{equation}
\label{entropyLovelock}
S=\frac{A}{4}\sum_ic_i\frac{i(n-2)!}{(n-2i)!}r_A^{2-2i}\, .
\end{equation}
The entropy (\ref{entropyLovelock}) is the same as that in the
static cases if one replaces  $r_{A} $ with the event horizon radius
$r_{+}$ of static black holes in Lovelock gravity~\cite{Cai4}.

Now, we have shown that $A\Psi^{(m)}$ term can be written as
$T\delta S$ term in the first law. Can the equations of motion in
Lovelock gravity be rewritten as the unified first law (\ref{unified
first law}) in Einstein gravity theory, where the left-hand side is
completely determined by spacetime geometry, while the right hand
side is determined by matter in spacetime? We can rewrite
(\ref{unified first law}) as
\begin{equation}
\label{rewriteEq} dE-A\Psi ^{(e)}-W^{(e)}dV=A\Psi ^{(m)}+W^{(m)}dV\,
.
\end{equation}
Note that the left hand side of the equation is totally determined
by geometry because  $\Psi^{(e)}$ and $W^{(e)}$ are defined by
geometric quantities as well as $E$. This implies that the left hand
side of the equation (\ref{rewriteEq}) can give us an
 energy form like the Misner-Sharp energy in Einstein gravity.
Indeed it is not hard to show that the left hand side of the
equation can be cast into a form of $dE_{L}$, where the function
$E_{L}$ is given by
\begin{equation}
\label{generalizedenergy}
E_{L}=\frac{\Omega _{n-2}}{16\pi }\sum_{i}^{p}\frac{(n-2)!c_{i}}{%
(n-2i-1)!}r^{n-2i-1}\left(1-f(v,r)\right)^{i}\, .
\end{equation}
Thus in the Vaidya-like spacetime for Lovelock gravity theory we
arrive at a generalized unified first law
\begin{equation}
dE_L=A\Psi ^{(m)}+W^{(m)}dV\, .
\end{equation}
After projecting on the apparent horizon, we obtain the first law of
apparent horizon
\begin{equation}
\label{physicalprocess} \delta E_{L}=T\delta S-P\delta V\, .
\end{equation}
where we have used  $W^{(m)}=-P$. If we write the function $f(v,r)$
back into the form $h^{ab}\partial_ar\partial_br$, the energy
(\ref{generalizedenergy}) is nothing but the generalized
Minser-Sharp energy suggested by H. Maeda and M. Nozawa in recent
papers~\cite{H.Maeda2, Nozawa:2007vq}, where the generalized
Misner-Sharp energy is shown to be a quasilocal conserved charge
associated with a locally conserved current constructed from the
generalized Kodama vector. Here let us stress that our procedure to
find the entropy and the energy expressions is a direct
generalization of the method proposed by Hayward to Lovelock gravity
theory. The two methods, though different, lead to the same result.
It would be interesting to use our method to study the unified first
law in other gravity theories and to get the corresponding
Miner-Sharp energy.

In some sense, Lovelock gravity theory is special since this theory
is ghost-free, and the equations of motion are in fact second order
for metric, as in the case of Einstein gravity. These properties may
help us separating matter from geometry effectively. For other
theories without such properties, such as $f(R)$ gravity, the
relation between equations of motion and thermodynamics is
complicated. Recently, Eling, Guedens and Jacobson have pointed out
that equations of motion for $f(R)$ gravity correspond to a
non-equilibrium thermodynamics of spacetime. To get correct
equations of motion, an entropy production term has to be added to
the Clausiu relation~\cite{Eling:2006aw}. However, to what extent
can the spacetime be described by equilibrium thermodynamics is
still an open question.

In summary, the definition of heat variation $\delta Q$ is important
in our procedure. The above discussion shows that $\delta Q$ should
be defined by pure matter energy supply or pure matter
energy-momentum tensor, through which we can find a reasonable
entropy associated with the apparent horizon. With this entropy and
the unified first law in Einstein gravity theory, we can obtain a
proper energy form, i.e. generalized Misner-Sharp energy, completely
determined by spacetime geometry. Thus we can establish the
generalized unified first law for Lovelock gravity theory.

The generalized unified first law (\ref{physicalprocess}) of the
apparent horizon is a version of physical process for the first law
of black hole thermodynamics, i.e., an active version of the first
law. In this version of the first law, one considers a spacetime and
matter energy-momentum tensor, and calculates the variation of
thermodynamical quantities which are induced by the matter (it
enters into horizon). The first law then establishes a relation
among the variations of those thermodynamic quantities.  In the
passive version or phase space version of first law, on the other
hand, one compares the thermodynamical quantities of two spacetimes
differing from each other by a small amount of variables, and then
gives variations of these quantities, which lead to the first law.
This physical process of the first law can shed some light on the
passive version of the first law. For the solution
(\ref{moregeneralsolution}), here we give some suggestion on the
passive version of the first law. By using equations
(\ref{moregeneralsolution}) and the definitions of the apparent
horizon and the surface gravity, the surface gravity can be written
as a function of the radius of the apparent horizon
\begin{equation}
\kappa =\frac{1}{2r_A \sum_{i} i \tilde{c}_i
r_A^{-2i+2}}\left[\sum_{i}(n-2i-1)\tilde{c}_ir_A^{-2i+2}+16\pi
\mathcal{C}(v)r_{A}^{(n-2)\sigma-n+4}\right]\, ,
\label{surfacegravity}
\end{equation}
where $r_A=r_A(v)$ and the coefficients $\tilde{c}_i$ are defined by
\begin{equation}
\tilde{c}_i=c_i\frac{(n-2)!}{(n-2i-1)!}\, .
\end{equation}
The energy inside the apparent horizon is given by
\begin{equation}
\label{bondilikemass} M_{L}=\frac{\Omega _{n-2}}{16\pi
}\sum_{i}^{p}\tilde{c}_ir_A^{n-2i-1}\, ,
\end{equation}
and the pressure on the apparent horizon by
\begin{equation}
P=-\mathcal{C}(v)r_A^{-(n-2)(1-\sigma)}\, .
\end{equation}
Then it is easy to check that the following relation holds
\begin{equation}
\label{phasespacefirstlaw} \mathrm{d} M_L=T\mathrm{d} S-P\mathrm{d}
V\, .
\end{equation}
where the entropy is given by (\ref{entropyLovelock}), while
$\mathrm{d} r_A$ denotes a variation of the apparent horizon, i.e.,
from $r_A$ to $r_A+\mathrm{d}r_A$. This is a variation of the state
parameter $r_A$ of the dynamical black hole. It should be noted that
the variation operator $``\mathrm{d}"$ is not an exterior derivative
$`` d "$ of the spacetime manifold, but a difference of two nearby
points in the solution space of the theory. Equation
(\ref{phasespacefirstlaw}) is just the passive version of the first
law. Another point we wish to stress is that the physical meaning of
energy $M_L$ is ambiguous. It is not an ADM mass of the spacetime
(although it is an ADM mass in the static limit), but a generalized
Misner-Sharp energy inside the apparent horizon. By using equation
(\ref{moregeneralsolution}), the energy $M_L$ can be expressed as
\begin{equation}
M_L=m(v)+\Omega_{n-2}\mathcal{C}(v)\Theta(r_A)\, .
\end{equation}
Thus, we see that the generalized Misner-sharp energy inside the
apparent horizon for the solution (\ref{moregeneralsolution}) gets
the physical meaning as a generalized Bondi mass if the matter is
pure radiation. However, for other type of matters, such as the term
leading to the nonvanishing $\mathcal{C}(v)$, the Bondi mass should
be modified. Further, it turns out difficult to define the Bondi
mass at null infinity in odd dimensions. This has been pointed out
in references~\cite{Hollands:2003ie, Hollands:2004ac}. For the
static case, $r_A=r_+$, equations (\ref{surfacegravity}) and
(\ref{phasespacefirstlaw}) reduce to the first law of black hole
thermodynamics in Lovelock gravity theory~\cite{Cai4}.

\section{conclusion and discussion}

In this paper, we have obtained a kind of generalized Vaidya
solutions (\ref{moregeneralsolution}) in a generic Lovelock gravity.
Explicit forms of these solutions are given for some special cases,
such as Gauss-Bonnet gravity,
 dimensional continued gravity and pure Lovelock gravity.  We have
also investigated the unified first law on the apparent horizon of
this spacetime, and given the  entropy (\ref{entropyLovelock})
associated with the apparent horizon and the energy
(\ref{generalizedenergy}) for $(n-2)$-dimensional sphere with radius
$r$. In our procedure, we treated all terms of Lovelock gravity
except for the Einstein tensor in equations of motion as an
effective energy-momentum tensor in Einstein gravity theory, to
which we could apply the unified first law proposed by Hayward for
Einstein gravity theory. Defining the heat flux by pure matter
energy-momentum tensor, we obtained an entropy expression
(\ref{entropyLovelock}) associated with the apparent horizon.
Substituting this entropy into the unified first law leads to an
energy form (\ref{generalizedenergy}) for Lovelock gravity, which is
the same as the one suggested by H. Maeda and M.
Nozawa~\cite{H.Maeda2,Nozawa:2007vq}. This method can be further
extended to other theories of gravity, for example, brane world
theory.

Here a comment not discussed in section III is in order. We did not
discuss whether the apparent horizon (or trapping horizon) is
``outer" and ``inner"~\cite{Hayward2}. In general, for higher
dimensional cases, the parameter $p$ in (\ref{1eq1}) is more than
one, so the equation (\ref{moregeneralsolution}) is a higher order
algebraic equation of $f(v,r)$ and may have multi-horizons. We have
assumed in section III that the apparent horizon is an outer
apparent horizon (trapping horizon) of black hole spacetime with
positive surface gravity $1/2 f' (v,r)|_{r = r_A}
>0$. In fact, the result of this paper also holds for the
cosmological event horizon (inner apparent horizon) with negative
surface gravity $1/2 f' (v,r)|_{r = r_A} <0$. In this case the
Hawking temperature of the apparent horizon is $T = |\kappa/2\pi|$.

Note that, while investigating the thermodynamical properties of the
new solution, we just focussed on the mechanics of the dynamical
black hole.  To  establish the thermodynamics of apparent horizon,
one has to show that there exists Hawking radiation with the
temperature $T=\kappa/2\pi$ as in the case of static black holes. In
fact, this may be proved by following some recent works on Hawking
radiation and entropy of dynamical
 spacetime~\cite{Viss,Di Criscienzo:2007fm, Hayward:2008jq, Kim:2008zt,
Cai:2008gw} (for a recent review, see~\cite{Nel}).

In a dynamical spacetime the apparent horizon is not necessarily the
same as the event horizon. For the solution
(\ref{moregeneralsolution}), the apparent horizon is simply given by
an algebraic equation $f(v,r_A(v))=0$, from which the event horizon
is very different. The event horizon requires the global information
of the spacetime. However, if we assume the event horizon is a null
hypersurface of the spacetime, it is not hard to find that the event
horizon of solution (\ref{moregeneralsolution}) is given by
$f(v,r_+(v))=2dr_+(v)/dv$. Therefore, in general, it is not easy to
give an explicit expression for the location of event horizon
because we have to solve a differential equation instead of an
algebraic equation. For some simple cases, such as the
four-dimensional Vaidya spacetime and the five-dimensional Vaidya
spacetime with the Gauss-Bonnet term, we may give the explicit
formula of the location of the event horizon for some simple mass
functions $m(v)$. Taking the four-dimensional Vaidya spacetime in
the Einstein gravity as an example, the event horizon is given by $
r_{+}=2m(v)/(1-2\dot{r}_{+})$. Some authors have calculated the
associated temperature and entropy by using brick wall model and
gravity anomalies method~\cite{zhao, Vagenas}:
\begin{equation}
T=\frac{1-2\overset{\cdot }{r}_{+}}{8\pi m(v)}=\frac{1}{4\pi r_{+}}\,,\quad
S=\frac{A}{4}=\pi r_{+}^{2}\,.
\end{equation}
At this stage a serious problem may arise. For a dynamical spacetime
such as the Vaidya solution, which horizon, the apparent horizon or
the event horizon or both of them, is indeed associated with Hawking
radiation and the entropy?  This issue has been commented in
\cite{Nel2} (references therein). But it is fair to say that this is
still an open question. Obviously, this is a quite interesting issue
worth further studying.

%%=============================================================

\section*{Acknowledgments}

RGC and YPH are supported partially by grants from NSFC, China (No.
10325525 and No. 90403029), and a grant from the Chinese Academy of
Sciences. LMC is supported by YST programs of APCTP, Korea. SPK is
supported by the Korea Research Council of Fundamental Science and
Technology (KRCF).

\end{document}